\documentstyle[12pt]{article}

\def\be{\begin{equation}}
\def\ee{\end{equation}}
\def\ba{\begin{eqnarray}}
\def\ea{\end{eqnarray}}

\def\ov{\overline}

\def\Ab{\bar{A}}

\def\E{{\hat{E}}}
\def\Pl{\ell_P}

\def\jd{j^{(d)}}
\def\ju{j^{(u)}}
\def\jdu{j^{(d+u)}}

\def\C*{$C^{\star}$}
\def\Cyl{{\rm Cyl}}

\def\ge{\geq}

\def\a{\alpha}

\def\e{\epsilon}
\def\g{\gamma}

\def\S{\Sigma}

\def\A{{\cal A}}
\def\ab{\ov{\cal A}}
\def\ag{{{\cal A}/{\cal G}}}
\def\agb{{\overline {{\cal A}/{\cal G}}}}

\def\C{{\cal C}}
\def\G{{\cal G}}
\def\Gb{{\overline \G}}
\def\H{{\cal H}}

\def\Ho{{\H}^o}

\def\Comp{{\mathchoice
{\setbox0=\hbox{$\displaystyle\rm C$}\hbox{\hbox to0pt
{\kern0.4\wd0\vrule height0.9\ht0\hss}\box0}}
{\setbox0=\hbox{$\textstyle\rm C$}\hbox{\hbox to0pt
{\kern0.4\wd0\vrule height0.9\ht0\hss}\box0}}
{\setbox0=\hbox{$\scriptstyle\rm C$}\hbox{\hbox to0pt
{\kern0.4\wd0\vrule height0.9\ht0\hss}\box0}}
{\setbox0=\hbox{$\scriptscriptstyle\rm C$}\hbox{\hbox to0pt
{\kern0.4\wd0\vrule height0.9\ht0\hss}\box0}}}}
\def\Co{{\mathchoice
{\setbox0=\hbox{$\displaystyle\rm C$}\hbox{\hbox to0pt
{\kern0.4\wd0\vrule height0.9\ht0\hss}\box0}}
{\setbox0=\hbox{$\textstyle\rm C$}\hbox{\hbox to0pt
{\kern0.4\wd0\vrule height0.9\ht0\hss}\box0}}
{\setbox0=\hbox{$\scriptstyle\rm C$}\hbox{\hbox to0pt
{\kern0.4\wd0\vrule height0.9\ht0\hss}\box0}}
{\setbox0=\hbox{$\scriptscriptstyle\rm C$}\hbox{\hbox to0pt
{\kern0.4\wd0\vrule height0.9\ht0\hss}\box0}}}}
\def\Rl{{\mathchoice
{\setbox0=\hbox{$\displaystyle\rm R$}\hbox{\hbox to0pt
{\kern0.4\wd0\vrule height0.9\ht0\hss}\box0}}
{\setbox0=\hbox{$\textstyle\rm R$}\hbox{\hbox to0pt
{\kern0.4\wd0\vrule height0.9\ht0\hss}\box0}}
{\setbox0=\hbox{$\scriptstyle\rm R$}\hbox{\hbox to0pt
{\kern0.4\wd0\vrule height0.9\ht0\hss}\box0}}
{\setbox0=\hbox{$\scriptscriptstyle\rm R$}\hbox{\hbox to0pt
{\kern0.4\wd0\vrule height0.9\ht0\hss}\box0}}}}

\let\ssection=\section 
\renewcommand{\section}{\setcounter{equation}{0}\ssection}

\begin{document}

\baselineskip=15pt

\vspace{.5in}

\centerline{{\Large {\bf Quantum Field Theory of Geometry}}
\footnote{Invited talk at the March '96 Boston Conference on 
Historical Examination and Philosophical Reflections On the 
Foundations of Quantum Field Theory (presented by AA).}}
\bigskip\bigskip
\centerline{Abhay Ashtekar${}^\star$ and Jerzy Lewanowski${}^\dagger$}
\medskip

\centerline{${}^\star$Center for Gravitational Physics and Geometry,}
\centerline{Physics Department, Penn State, University Park,
PA 16802-6300, U.S.A.}
\centerline{${}^\dagger$Institute of Theoretical Physics}
\centerline{University of Warsaw, ul Hoza 69, 00-681 Warsaw, Poland}
\centerline{${}^\dagger$Max Planck Institut f\"ur Gravitationphysik}
\centerline{Schlaatzweg 1, 14473 Potsdam, Germany}
\bigskip
\bigskip


\pagenumbering{arabic}

\begin{section} {Introduction}
\label{s1}

Several speakers at this conference have emphasized the conceptual
difficulties of quantum gravity (see particularly \cite{aw,bd,cr}).
As they pointed out, when we bring in gravity, some of the basic
premises of quantum field theory have to undergo radical changes: {\it
we must learn to do physics in absence of a background space-time
geometry}.  This immediately leads to a host of technical difficulties
as well. For, the familiar mathematical methods of quantum field
theory are deeply rooted in the availability of a fixed space-time
metric, which, furthermore is generally taken to be flat. The purpose
of this contribution is to illustrate how these conceptual and
technical difficulties can be overcome.

For concreteness, we will use a specific non-perturbative approach
and, furthermore, limit ourselves to just one set of issues: Exploration
of the nature of quantum geometry. Nonetheless, the final results have
a certain degree of robustness and the constructions involved provide
concrete examples of ways in which one can analyze genuine field
theories, with an infinite number of degrees of freedom, in absence of
a background metric. As we will see, the underlying diffeomorphism
invariance is both a curse and a blessing. On the one hand, since
there is so little background structure, concrete calculations are
harder and one is forced to invent new regularization methods. On the
other hand, when one does succeed, the final form of results is often
remarkably simple since the requirement of diffeomorphism invariance
tends to restrict the answers severely. The final results are often
unexpected and qualitatively different from the familiar ones
from standard quantum field theories.

Let us begin with a brief discussion of the issue on which we wish to
focus.  In his celebrated inaugural address, Riemann suggested
\cite{Rie} that geometry of space may be more than just a fiducial,
mathematical entity serving as a passive stage for physical phenomena,
and may in fact have direct physical meaning in its own right. As we
know, general relativity provided a brilliant confirmation of this
vision: Einstein's equations put geometry on the same footing as
matter. Now, the physics of this century has shown us that matter has
constituents and the 3-dimensional objects we perceive as solids in
fact have a discrete underlying structure. The continuum description
of matter is an approximation which succeeds brilliantly in the
macroscopic regime but fails hopelessly at the atomic scale. It is
therefore natural to ask if the same is true of geometry. Does
geometry also have constituents at the Planck scale?  What are its
atoms? Its elementary excitations?  Is the space-time continuum only a
`coarse-grained' approximation?  If so, what is the nature of the
underlying quantum geometry?

To probe such issues, it is natural to look for hints in the
procedures that have been successful in describing matter. Let us
begin by asking what we mean by quantization of physical quantities.
Take a simple example --the hydrogen atom. In this case, the answer is
clear: while the basic observables --energy and angular momentum--
take on a continuous range of values classically, in quantum mechanics
their spectra are discrete. So, we can ask if the same is true of
geometry. Classical geometrical observables such as areas of surfaces
and volumes of regions can take on continuous values on the phase
space of general relativity. Are the spectra of corresponding quantum
operators discrete? If so, we would say that geometry is quantized.

Thus, it is rather easy to pose the basic questions in a precise
fashion. Indeed, they could have been formulated soon after the advent
of quantum mechanics. Answering them, on the other hand, has proved to
be surprisingly difficult. The main reason, we believe, is the
inadequacy of the standard techniques. More precisely, in the
traditional approaches to quantum field theory, one {\it begins} with
a continuum, background geometry. To probe the nature of quantum
geometry, on the other hand, we should not begin by assuming the
validity of this model. We must let quantum gravity decide whether
this picture is adequate at the Planck scale; the theory itself should
lead us to the correct microscopic model of geometry.

With this general philosophy, in this talk we will use a
non-perturbative, canonical approach to quantum gravity to probe the
nature of quantum geometry. In this approach, one uses $SU(2)$
connections on a 3-manifold as configuration variables; 3-metrics are
constructed from `electric fields' which serve as the conjugate
momenta. These are all dynamical variables; to begin with, we are
given only a 3-manifold without {\it any} fields.  Over the past three
years, this approach has been put on a firm mathematical footing
through the development of a new functional calculus on the space of
gauge equivalent connections [4-12]. This calculus does not use any
background fields (such as a metric) and is therefore well-suited for
a fully non-perturbative exploration of the nature of quantum
geometry. 

In section 2, we will introduce the basic tools from this functional
calculus and outline our general strategy. This material is then used
in section 3 to discuss the main results. In particular, operators
corresponding to areas of 2-surfaces and volumes of 3-dimensional
regions are regulated in a fashion that respects the underlying
diffeomorphism invariance. They turn out to be self-adjoint on the
underlying (kinematical) Hilbert space of states. A striking property
is that their spectra are {\it purely} discrete. This indicates that
the underlying quantum geometry is far from what the continuum picture
might suggest. Indeed, the fundamental excitations of quantum geometry
are 1-dimensional, rather like polymers, and the 3-dimensional
continuum geometry emerges only on coarse graining \cite{9,19}. In the
case of the area operators, the spectrum is explicitly known.  This
detailed result should have implications on the statistical mechanical
origin of the black hole entropy \cite{bm,car} and the issue is being
investigated. Section 4 discusses a few ramifications of the main
results.

Our framework belongs to what Carlo Rovelli referred to in his talk as
`general quantum field theory'. Thus, our constructions do not
directly fall in the category of axiomatic or constructive quantum
field theory and, by and large, our calculations do not use the
standard methods of perturbative quantum field theory. Nonetheless, we
{\it do} discuss the quantum theory of a system with an infinite
number of degrees of freedom (which, moreover, is diffeomorphism
covariant) and face the issues of regularization squarely. For this,
we begin `ab-initio', construct the Hilbert space of states, introduce
on it well-defined operators which represent (regulated) geometric
observables and examine their properties.

Details of the results discussed here can be found in \cite{al5,al6,
jl}.  At a conceptual level, there is a close similarity between the
basic ideas used here and those used in discussions based on the `loop
representation' \cite{9,10, rl1, rl2}. (For a comparison, see
\cite{al5, jl}). Indeed, the development of the functional calculus
which underlies this analysis was itself motivated, in a large
measure, by the pioneering work on loop representation by Rovelli and
Smolin \cite{11}. Finally, we emphasize that this is {\it not} a
comprehensive survey of non-perturbative quantum gravity; our main
purpose, as mentioned already, is to illustrate how one can do quantum
field theory in absence of a space-time background and to point out
that results can be unexpected. Indeed, even the use of general
relativity as the point of departure is only for concreteness; the
main results do not depend on the details of Einstein's equations.%
\footnote{Nonetheless, since there were several remarks in this
conference on the viability of quantum general relativity, it is
appropriate to make a small digression to clarify the situation.  It
is well-known that general relativity is perturbatively
non-renormalizable. However, there {\it do} exist quantum field
theories which share this feature with general relativity but are {\it
exactly soluble}.  A striking example is ${\rm (GN)}_3$, the
Gross-Neveau model in 3 dimensions. Furthermore, in the case of
general relativity, there are {\it physical} reasons which make
perturbative methods especially unsuitable. Whether quantum general
relativity can exist non-perturbatively is, however, an open
question. For further details and current status, see
e.g. \cite{aa}.)}  

\end{section}

\begin{section}{Tools}
\label{s2}

This section is divided in to four parts. The first summarizes the
formulation of general relativity based on connections; the second
introduces the quantum configuration space; the third presents an
intuitive picture of the {\it non-perturbative} quantum states and the
fourth outlines our strategy to probe quantum geometry.

\begin{subsection}{From metrics to connections}
\label{s2.1}

The non-perturbative approach we wish to use here has its roots in
canonical quantization.  The canonical formulation of general
relativity was developed in the late fifties and early sixties in a
series of papers by Bergmann, Dirac and Arnowitt, Deser and Misner.
In this formulation, general relativity arises as a dynamical theory
of 3-metrics. The framework was therefore named {\it geometrodynamics}
by Wheeler and used as a basis for canonical quantization both by him
and his associates and by Bergmann and his collaborators. The
framework of geometrodynamics has the advantage that classical
relativists have a great deal of geometrical intuition and physical
insight into the nature of the basic variables --3-metrics $g_{ab}$
and extrinsic curvatures $K_{ab}$.  For these reasons, the framework
has played a dominant role, e.g., in numerical relativity. However, it
also has two important drawbacks.  First, it sets the mathematical
treatment of general relativity quite far from that of theories of
other interactions where the basic dynamical variables are connections
rather than metrics. Second, the equations of the theory are rather
complicated in terms of metrics and extrinsic curvatures; being
non-polynomial, they are difficult to carry over to quantum theory
with a reasonable degree of mathematical precision.

For example, consider the standard Wheeler-DeWitt equation:
\be 
\big[{G\hbar\over \sqrt{g}}
(g^{ab} g^{cd} - {1\over 2} g^{ac}g^{bd})\,\, 
{\delta\over{\delta g_{ac}}} {\delta\over{\delta g_{bd}}}\, - 
\, {\sqrt{g}\over{G\hbar}}\, R(g) \big]\circ 
\Psi (g) = 0\,  
\ee
where $g$ is the determinant of the 3-metric $g_{ab}$ and $R$ its
scalar curvature.  As is often emphasized, since the kinetic term
involves products of functional derivatives evaluated at the same
point, it is ill-defined.  However, there are also other, deeper
problems. These arise because, in field theory, the quantum
configuration space --the domain space of wave functions $\Psi$-- is
larger than the classical configuration space. While we can restrict
ourselves to suitably smooth fields in the classical theory, in
quantum field theory, we are forced to allow distributional field
configurations. Indeed, even in the free field theories in Minkowski
space, the Gaussian measure that provides the inner product is
concentrated on genuine distributions. This is the reason why in
quantum theory fields arise as operator-valued distributions.  One
would expect that the situation would be at least as bad in quantum
gravity. If so, even the products of the 3-metrics that appear in
front of the momenta as well as the scalar curvature in the potential
term would fail to be meaningful. Thus, the left hand side of the
Wheeler-DeWitt equation is seriously ill-defined and must be
regularized appropriately.

However, as we just said, the problem of distributional configurations
arises already in the free field theory in Minkowski
space-time. There, we do know how to regularize physically interesting
operators. So, why can we not just apply those techniques in the
present context? The problem is that those techniques are tied to the
presence of a background Minkowski metric. The covariance of the
Gaussian measure, for example, is constructed from the Laplacian
operator on a space-like plane defined by the induced metric and
normal ordering and point-splitting regularizations also make use of
the background geometry. In the present case, we do {\it not} have
background fields at our disposal. We therefore need to find another
avenue. What is needed is a suitable functional calculus --integral
and differential-- that respects the diffeomorphism invariance of the
theory.

What space are we to develop this functional calculus on? Recall first
that, in the canonical approach to diffeomorphism invariant theories
such as general relativity or supergravity, the key mathematical
problem is that of formulating and solving the quantum constraints.
(In Minkowskian quantum field theories, the analogous problem is that
of defining the regularized quantum Hamiltonian operator.) It is
therefore natural to work with variables which, in the classical
theory, simplify the form of the constraints. It turns out that, from
this perspective, connections are better suited than metrics\cite{20}.

We will conclude by providing explicit expressions of these
connections. Recall first that in geometrodynamics we can choose as
our canonical pair, the fields $(E^a_i, K_a^i)$ where $E^a_i$ is
a triad (with density weight one) and $K_a^i$, the extrinsic
curvature.  Here $a$ refers to the tangent space of the 3-manifold and
$i$ is the internal $SO(3)$ --or, $SU(2)$, if we wish to consider
spinorial matter-- index. The triad is the square-root of the metric
in the sense that $E^a_i E^{bi} =: g g^{ab}$, where $g$ is the
determinant of the covariant 3-metric $g_{ab}$, and $K_a^i$ is related
to the extrinsic curvature $K_{ab}$ via: $K_a^i=
(1/\sqrt{g})K_{ab}E^{bi}$.  Let us make a change of variables: 
\be
(E^a_i, K_a^i)\,\, \mapsto \,\, (A_a^i:= \Gamma_a^i - 
K_a^i, \, E^a_i),
\ee
where $\Gamma_a^i$ is the spin connection determined by the triad.  It
is not difficult to check that this is a canonical transformation on
the real phase space \cite{20,21}. It will be convenient to regard
$A_a^i$ as the configuration variable and $E^a_i$ as the conjugate
momentum so that the phase space has the same structure as in the
$SU(2)$ Yang-Mills theory. The basic Poisson bracket relations are:
\be\label{pb} 
\{A_a^i(x),\, E^b_j(y)\} = G\, \delta_a^b \delta_j^i \delta^3 (x,y)\, ,
\ee
where the gravitational constant, $G$, features in the Poisson bracket
relations because $E^a_i(x)$ now has the physical dimensions of a
triad rather than that of Yang-Mills electric field. In terms of these
variables, general relativity has the same kinematics as Yang-Mills
theory. Indeed, one of the constraints of general relativity is
precisely the Gauss constraint of Yang-Mills theory. Thus, the phase
spaces of the two theories are the same and the constraint surface of
general relativity is embedded in that of Yang-Mills
theory. Furthermore, in terms of these variables, the remaining
constraints of general relativity simplify considerably. Indeed, there
is a precise sense in which they are the simplest non-trivial
equations one can write down in terms of $A_a^i$ and $E^a_i$ without
reference to any background field \cite{14}. Finally, (in the
spatially compact context) the Hamiltonian of general relativity is
just a linear combination of constraints.

To summarize, one can regard the space $\ag$ of $SU(2)$ connections
modulo gauge transformations on a (`spatial') 3-manifold $\S$ as the
classical configuration space of general relativity.
\end{subsection}

\begin{subsection} {Quantum configuration space}
\label{s2.2}

As we already indicated, in Minkowskian quantum field theories, the
quantum configuration space includes distributional fields which are
absent in the classical theory and physically interesting measures are
typically concentrated on these `genuinely quantum' configurations.
The overall situation is the same in general relativity.

Thus, the quantum configuration space $\agb$ is a certain completion
of $\ag$ \cite{1,2}.  $\agb$ inherits the quotient structure of $\ag$,
i.e., $\agb$ is the quotient of the space $\ab$ of generalized
connections by the space $\Gb$ of generalized gauge
transformations. To see the nature of the generalization involved,
recall first that each smooth connection defines a holonomy along
paths%
\footnote{For technical reasons, we will assume that all paths are
analytic. An extension of the framework to allow for smooth paths is
being carried out \cite{15}. The general expectation is that the main
results will admit natural generalizations to the smooth category.} %
in $\S$: $h_p(A):= {\cal P}\exp -\int_p A$. Generalized connections
capture this notion. That is, each $\Ab$ in $\ab$ can be defined
\cite{3,5} as a map which assigns to each oriented path $p$ in $\S$ an
element $\Ab(p)$ of $SU(2)$ such that: i) $\Ab(p^{-1}) =
(\Ab(p))^{-1}$; and, ii) $\Ab(p_2\circ p_1) = \Ab(p_2)\cdot\Ab(p_1)$,
where $p^{-1}$ is obtained from $p$ by simply reversing the
orientation, $p_2\circ p_1$ denotes the composition of the two paths
(obtained by connecting the end of $p_1$ with the beginning of $p_2$)
and $\Ab(p_2)\cdot\Ab(p_1)$ is the composition in $SU(2)$. A
generalized gauge transformation is a map $g$ which assigns to each
point $v$ of $\S$ an $SU(2)$ element $g(x)$ (in an arbitrary, possibly
discontinuous fashion). It acts on $\Ab$ in the expected manner, at
the end points of paths: $\Ab(p)\rightarrow g(v_+)^{-1} \cdot\Ab(p)\cdot
g(v_-)$, where $v_{-}$ and $v_+$ are respectively the beginning
and the end point of $p$. If $\Ab$ happens to be a smooth connections,
say $A$, we have $\Ab(p) = h_p(A)$.  However, in general, $\Ab(p)$ can
not be expressed as a path ordered exponential of a smooth 1-form with
values in the Lie algebra of $SU(2)$ \cite{2}. Similarly, in general,
a generalized gauge transformation can not be represented by a smooth
group valued function on $\S$.

At first sight the spaces $\ab$, $\Gb$ and $\agb$ seem too large to be
mathematically controllable. However, they admit three
characterizations which enables one to introduce differential and
integral calculus on them \cite{1,2,4}. We will conclude this
sub-section by summarizing the characterization --as suitable limits
of the corresponding spaces in lattice gauge theory-- which will be
most useful for the main body of this paper.

Fix a graph $\g$ in the 3-manifold $\S$. In the physics terminology,
one can think of a graph as a `floating lattice', i.e., a lattice
whose edges are not required to be rectangular. (Indeed, they may even
be non-trivially knotted!) Using the standard ideas from lattice gauge
theory, we can construct the configuration space associated with the
graph $\g$. Thus, we have the space $\A_\g$, each element $A_\g$ of
which assigns to every edge in $\g$ an element of $SU(2)$ and the
space $\G_\g$ each element $g_\g$ of which assigns to each vertex in
$\g$ an element of $SU(2)$. (Thus, if $N$ is the number of edges in
$\g$ and $V$ the number of vertices, $\A_\g$ is isomorphic with
$[SU(2)]^N$ and $\G_\g$ with $[SU(2)]^V$). $\G_\g$ has the obvious
action on $\A_\g$: $A_\g(e) \rightarrow g(v_+)^{-1}\cdot A_\g(e)\cdot
g(v_-)$.  The (gauge invariant) configuration space associated with
the floating lattice $\gamma$ is just $\A_\g/\G_\g$. The spaces $\ab$,
$\Gb$ and $\agb$ can be obtained as well-defined (projective) limits
of the spaces $\A_\g$, $\G_\g$ and $\A_\g/\G_\g$ \cite{4,2}. Note
however that this limit is {\it not} the usual `continuum limit' of a
lattice gauge theory in which one lets the edge length go to
zero. Here, we are already in the continuum and have available to us
{\it all possible} floating lattices from the beginning. We are just
expressing the quantum configuration space of the continuum theory as
a suitable limit of the configuration spaces of theories associated
with all these lattices.

To summarize, the quantum configuration space $\agb$ is a specific
extension of the classical configuration space $\ag$. Quantum states
can be expressed as complex-valued, square-integrable functions on
$\agb$, or, equivalently, as $\Gb$-invariant square-integrable
functions on $\ab$.  As in Minkowskian field theories, while $\ag$ is
dense in $\agb$ topologically, measure theoretically it is generally
sparse; typically, $\ag$ is contained in a subset set of zero measure
of $\agb$ \cite{4}.  Consequently, what matters is the value of wave
functions on `genuinely' generalized connections.  In contrast with
the usual Minkowskian situation, however, $\ab$, $\Gb$ and $\agb$ are
all {\it compact} spaces in their natural (Gel'fand) topologies
[4-8]. This fact simplifies a number of technical issues.

\end{subsection}

\begin{subsection} {Hilbert space}
\label{s2.3}

Since $\agb$ is compact, it admits regular (Borel, normalized)
measures and for every such measure we can construct a Hilbert space
of square-integrable functions.  Thus, to construct the Hilbert space
of quantum states, we need to select a specific measure on $\agb$.

It turns out that $\ab$ admits a measure $\mu^o$ that is preferred by
both mathematical and physical considerations \cite{2,3}.
Mathematically, the measure $\mu^o$ is natural because its definition
does not involve introduction of any additional structure: it is
induced on $\ab$ by the Haar measure on $SU(2)$. More precisely, since
$\A_\g$ is isomorphic to $[SU(2)]^N$, the Haar measure on $SU(2)$
induces on it a measure $\mu^o_\g$ in the obvious fashion.  As we vary
$\g$, we obtain a family of measures which turn out to be compatible
in an appropriate sense and therefore induce a measure $\mu^o$ on
$\ab$.  This measure has the following attractive properties \cite{2}:
i) it is faithful; i.e., for any continuous, non-negative function $f$
on $\ab$, $\int d\mu^o \, f \ge 0$, equality holding if and only if
$f$ is identically zero; and, ii) it is invariant under the (induced)
action of ${\rm Diff}[\S]$, the diffeomorphism group of $\S$.
Finally, $\mu^o$ induces a natural measure $\tilde{\mu}^o$ on $\agb$:
$\tilde{\mu}^o$ is simply the push-forward of $\mu^o$ under the
projection map that sends $\ab$ to $\agb$.  Physically, the measure
$\tilde{\mu}^o$ is selected by the so-called `reality conditions'.
More precisely, the classical phase space admits an (over)complete set
of naturally defined configuration and momentum variables which are
real, and the requirement that the corresponding operators on the
quantum Hilbert space be self-adjoint selects for us the measure
$\tilde{\mu}^o$ \cite{7}. 

Thus, it is natural to use $\tilde\Ho := L^2(\agb, d\tilde{\mu}^o)$ as
our Hilbert space. Elements of $\tilde\Ho$ are the kinematic states;
we are yet to impose quantum constraints. Thus, $\tilde\Ho$ is the
classical analog of the {\it full} phase-space of quantum gravity
(prior to the introduction of the constraint sub-manifold). Note that
these quantum states can be regarded also as {\it gauge invariant}
functions on $\ab$. In fact, since the spaces under consideration are
compact and measures normalized, we can regard $\tilde\Ho$ as the
gauge invariant {\it sub-space} of the Hilbert space $\Ho := L^2(\ab,
d\mu^o)$ of square-integrable functions on $\ab$ \cite{3,4}. {\em In
what follows, we we will often do so}.

What do `typical' quantum states look like? To provide an intuitive
picture, we can proceed as follows. Fix a graph $\g$ with $N$ edges and
consider functions $\Psi_\g$ of generalized connections of the form
$\Psi_\g (\Ab) = \psi (\Ab(e_1),..., \Ab(e_N))$ for {\it some} smooth
function $\psi$ on $[SU(2)]^N$, where $e_1, ..., e_N$ are the edges of
the graph $\g$.  Thus, the functions $\Psi_\g$ know about what the
generalized connections do only to those paths which constitute the
edges of the graph $\g$; they are precisely the quantum states of the
gauge theory associated with the `floating lattice' $\g$. This space
of states, although infinite dimensional, is quite `small' in the
sense that it corresponds to the Hilbert space associated with a
system with only a {\it finite} number of degrees of freedom. However,
if we vary $\g$ through all possible graphs, the collection of all
states that results is very large.  Indeed, one can show that it is
{\it dense} in the Hilbert space $\Ho$. (If we restrict ourselves to
$\Psi_\g$ which are gauge invariant, we obtain a dense sub-space in
$\tilde\Ho$.) 

Since each of these states $\Psi_\g$ depends only on a finite number
of variables, borrowing the terminology from the quantum theory of
free fields in Minkowski space, they are called {\it cylindrical
functions} and denoted by $\Cyl$. Gauge invariant cylindrical
functions represent the `typical' kinematic states.  In many ways,
$\Cyl$ is analogous to the space $C_o^\infty(R^3)$ of smooth functions
of compact support on $R^3$ which is dense in the Hilbert space
$L^2(R^3, d^3x)$ of quantum mechanics. Just as one often defines
quantum operators -- e.g., the position, the momentum and the
Hamiltonians-- on $C^\infty_o$ first and then extends them to an
appropriately larger domain in the Hilbert space $L^2(R^3, d^3x)$, we
will define our operators first on $\Cyl$ and then extend them
appropriately.

Cylindrical functions provide considerable intuition about the nature
of quantum states we are led to consider. These states represent
1-dimensional polymer-like excitations of geometry/gravity rather than
3-dimensional wavy undulations on flat space. Just as a polymer,
although intrinsically 1-dimensional, exhibits 3-dimensional
properties in sufficiently complex and densely packed configurations,
the fundamental 1-dimensional excitations of geometry can be packed
appropriately to provide a geometry which, when coarse-grained on scales
much larger than the Planck length, lead us to continuum geometries
\cite{9,18}.  Thus, in this description, gravitons can arise only as
approximate notions in the low energy regime \cite{19}.  At the basic
level, states in $\tilde\Ho$ are fundamentally different from the Fock
states of Minkowskian quantum field theories. The main reason is the
underlying diffeomorphism invariance: In absence of a background
geometry, it is not possible to introduce the familiar Gaussian
measures and associated Fock spaces.

\end{subsection}

\begin{subsection}{Statement of the problem}
\label{s2.4}

We can now outline the general strategy that will be followed in
section 3.

Recall that the classical configuration variable is an $SU(2)$
connection%
\footnote{We assume that the underlying 3-manifold $\S$ is orientable.
Hence, principal $SU(2)$ bundles over $\S$ are all topologically
trivial. Therefore, we can represent the $SU(2)$ connections on the
bundle by a $su(2)$-valued 1-form on $\S$. The matrices $\tau_i$ are
anti-Hermitian, given, e.g., by $(-i/2)$-times the Pauli matrices.} %
$A_a^i$ on a 3-manifold $\S$, where $i$ is the $su(2)$-internal index
with respect to a basis $\tau_i$. Its conjugate momentum $E^b_j$ has
the geometrical interpretation of an orthonormal triad with density
weight one \cite{20}.  Therefore, geometrical observables
--functionals of the 3-metric-- can be expressed in terms of this
field $E^a_i$. Fix within the 3-manifold $\S$ any analytic, finite
2-surface. The area $A_S$ of $S$ is a well-defined, real-valued
function on the {\it full} phase space of general relativity (which
happens to depend only on $E^a_i$).  It is easy to verify that these
kinematical observables can be expressed as:
\be\label{2.4} 
A_S := \int_S dx^1\wedge dx^2 \, [E^3_i E^{3i}]^{1\over 2}\, ,
\ee
where, for simplicity, we have used adapted coordinates such that $S$
is given by $x^3 = 0$, and $x^1,x^2$ parameterize $S$, and where the
internal index $i$ is raised by a the inner product we use on $su(2)$,
$k(\tau_i,\tau_j) = -2{\rm Tr}(\tau_i\tau_j)$. Similarly, if $R$ is
any 3-dimensional open region within $\S$, the associated 
volume is a function on the phase space given by:
\be
\label{v} V_R := \int_R dx^1\wedge dx^2\wedge dx^3\, [{1\over
3}\eta_{abc}\e^{ijk} E^a_i E^b_j E^c_k]^{1\over 2}\, , 
\ee
where $\eta_{abc}$ is the (metric independent, natural) Levi-Civita
density of weight $-1$ on $\S$.  Our task is to find the corresponding
operators on the kinematical Hilbert space $\tilde\Ho$ and investigate
their properties.

There are several factors that make this task difficult.  Intuitively,
one would expect that $E^a_i(x)$ to be replaced by the
`operator-valued distribution' $-i\hbar G\delta/\delta A_a^i(x)$.
(See the basic Poisson bracket relation (\ref{pb}). Unfortunately,
the classical expression of $A_S$ involves {\it square-roots of
products} of $E$'s and hence the formal expression of the
corresponding operator is badly divergent.  One must introduce a
suitable regularization scheme.  However, we do not have at our
disposal the usual machinery of Minkowskian field theories and even
the precise rules that are to underlie such a regularization are not
apriori clear.

There are however certain basic expectations that we can use as
guidelines: i) the resulting operators should be well-defined on a
dense sub-space of $\tilde\Ho$; ii) their final expressions should be
diffeomorphism covariant, and hence, in particular, independent of any
background fields that may be used in the intermediate steps of the
regularization procedure; and, iii) since the classical observables
are real-valued, the operators should be self-adjoint. These
expectations seem to be formidable at first. Indeed, these demands are
rarely met even in Minkowskian field theories; in presence of
interactions, it is extremely difficult to establish rigorously that
physically interesting operators are well-defined and self-adjoint. As
we will see, the reason why one can succeed in the present case is
two-folds. First, the requirement of diffeomorphism covariance is a
powerful restriction that severely limits the possibilities. Second,
the background independent functional calculus is extremely
well-suited for the problem and enables one to circumvent the various
road blocks in subtle ways.

Our general strategy will be following. We will define the regulated
versions of area and volume operators on the dense sub-space $\Cyl$ of
cylindrical functions and show that they are essentially self-adjoint
(i.e., admit unique self-adjoint extensions to $\tilde\Ho$).  This
task is further simplified because the operators leave each sub-space
$\H_\g$ spanned by cylindrical functions associated with any one graph
$\g$ invariant. This in effect reduces the field theory problem (i.e.,
one with an infinite number of degrees of freedom) to a quantum
mechanics problem (in which there are only a finite number of degrees
of freedom). Finally, the operators in fact leave invariant certain
{\it finite} dimensional sub-space of $\Ho$ (associated with `extended
spin networks' \cite{al5}).  This powerful simplification further
reduces the task of investigating the properties of these operators;
in effect, the quantum mechanical problem (in which the Hilbert space
is still infinite dimensional) is further simplified to a problem
involving spin systems (where the Hilbert space is finite
dimensional). It is because of these simplifications that a detailed
analysis becomes possible.

\end{subsection}
\end{section}

\begin{section}{Quantum Geometry}
\label{s3}

Our task is to construct a well-defined operator $\hat{A}_S$ and
$\hat{V}_R$ starting from the classical expressions (\ref{2.4},
\ref{v}). As is usual in quantum field theory, we will begin with the
formal expression obtained by replacing $E^a_i$ in (\ref{2.4},\ref{v})
by the corresponding operator valued distribution $\E^a_i$ and then
regulate it to obtain the required operators. (For an early
discussion of non-perturbative regularization, see, in particular,
\cite{bp}). For brevity, we will discuss the area operators in some detail
and then give the final result for the volume operators. Furthermore,
to simplify the presentation, we will assume that $S$ is covered
by a single chart of adapted coordinates. Extension to the general
case is straightforward: one mimics the procedure used to define the
integral of a differential form over a manifold. That is, one takes
advantage of the coordinates invariance of the the resulting `local'
operator and uses a partition of unity.

\begin{subsection}{regularization}
\label{s3.1}
                        
The first step in the regularization procedure is to smear (the
operator analog of) $E^3_i(x)$ and point split the integrand in
(\ref{2.4}). Since in this integrand the point $x$ lies on the
2-surface $S$, let us try to use a 2-dimensional smearing
function. Let $f_\epsilon (x, y)$ be a 1-parameter family of fields on
$S$ which tend to the $\delta(x,y)$ as $\epsilon$ tends to zero; i.e.,
such that 
\be 
\lim_{\epsilon\rightarrow 0} \int_S d^2y \,
f_\epsilon(x^1,x^2; y^1,y^2) g(y^1,y^2) = g(x^1,x^2)\, , 
\ee 
for all smooth densities $g$ of weight $1$ and of compact support on
$S$. (Thus, $f_\e(x,y)$ is a density of weight 1 in $x$ and a function
in $y$.) The smeared version of $E^3_i(x)$ will be defined to be: 
\be 
[E^3_i]_f (x):= \int_S d^2y\, f_\epsilon(x,y) E^3_i(y)\, , 
\ee 
so that, as $\epsilon$ tends to zero, $[E^3_i]_f$ tends to
$E^3_i(x)$. The point-splitting strategy now provides a `regularized
expression' of area: 
\ba\label{ra} [A_S]_f &:=& \int_S d^2x \,
\big[\int_S d^2y, f_\e (x,y) E^3_i(y)\, \int_S d^2z \, f_\e (x,z)
E^{3i}(z)\, \big]^{1\over 2}\nonumber \\ &=& \int_S d^2x\,
\big[[E^3_i]_f (x) [E^{3i}]_f (x) \big]^{1\over 2}\, , 
\ea 
which will serve as the point of departure in the subsequent
discussion.  To simplify technicalities, we will assume that the
smearing field $f_\e(x,y)$ has the following additional properties for
sufficiently small $\e > 0$: i) for any given $y$, $f_\e (x,y)$ has
compact support in $x$ which shrinks uniformly to $y$; and, ii)
$f_\e(x,y)$ is non-negative. These conditions are very mild and we are
thus left with a large class of regulators.%
\footnote{For example, $f_\e(x,y)$ can be constructed as follows. Take
{\it any} non-negative function $f$ of compact support on $S$ such
that $\int d^2x f(x) = 1$ and set $f_\e(x,y) = (1/\e^2)f((x-y)/\e)$.
Here, we have used the given chart to write $x-y$ and give $f_\e(x,y)$
a density weight in $x$.}

First, let us fix a graph $\g$ and consider a cylindrical function
$\Psi_\g$ on $\ab$, 
\be\label{cyl} \Psi_\g(\Ab) = \psi (\Ab(e_1),
.., \Ab(e_N))\, \equiv \psi(g_1, ..., g_n) 
\ee 
where, as before, $N$ is the total number of edges of $\g$, $g_k =
\Ab(e_k)$ and where $\psi$ is a smooth function on $[SU(2)]^N$. One
can show \cite{al5} that the action of the regulated triad operator on
such a state is given by:
\be [\E^3_i]_f (x) \,\cdot\, \Psi_\g =\ {i\Pl^2\over 2}
\Big[\sum_{I=1}^{N} \kappa_I\, f_\e (x,v_{\alpha_I})\,\,
X_I^i\Big]\,\cdot\, \psi(g_1, ..., g_N)\, 
\ee 
Here, $X_I^i$ are the left/right invariant vector fields on the $I$th
group copy in the argument of $\psi$ in the $i$th internal direction,
i.e., are operators assigned to the edge $e_I$ by the following
formula
\be\label{X} 
X_I^i\,\cdot\, \psi(g_1, ... ,g_N) = \cases{
(g_I\tau^i)^A_B\, {\partial\psi\over \partial (g_I)^A_B},&
when $e_I$ is outgoing\cr 
-(\tau^i g_I)^A_B\, {\partial\psi\over
\partial (g_I)^A_B},& when $e_I$ is incoming,\cr} 
\ee 
and $\kappa_I$ are real numbers given by: 
\be \kappa_I= \cases{\,0,& if
$e_I$ is tangential to $S$ or does not intersect $S$, \cr
       +1,& if $e_I$ has an isolated intersection with $S$ and 
        lies above $S$ \cr 
      -1,& if $e_I$ has an isolated intersection with $S$ and 
       lies below $S$.\cr} 
\ee

The right side again defines a cylindrical function based on the
(same) graph $\g$. Denote by $\Ho_\g$ the Hilbert space $L^2(\A_\g,
d\mu^o_\g)$ of square integrable cylindrical functions associated with
a fixed graph $\g$. Since $\mu^o_\g$ is the induced Haar measure on
$\A_\g$ and since the operator is just a sum of right/left invariant
vector fields, standard results in analysis imply that, with domain
${\rm Cyl}_\g^1$ of all $C^1$ cylindrical functions based on $\g$, it
is an essentially self-adjoint on $\Ho_\g$. Now, it is straightforward
to verify that the operators on $\Ho_\g$ obtained by varying $\g$ are
all compatible%
\footnote{Given two graphs, $\g$ and $\g'$, we say that $\g\ge \g'$ if
and only if every edge of $\g'$ can be written as a composition of
edges of $\g$. Given two such graphs, there is a projection map from
$\A_\g$ to $\A_{\g'}$, which, via pull-back, provides an unitary
embedding $U_{\g,\g'}$ of $\tilde\Ho_{\g'}$ into $\tilde\Ho_\g$. A
family of operators ${\cal O}_\g$ on the Hilbert spaces $Ho_\g$ is
said to be compatible if $U_{\g,\g'}{\cal O}_{\g'} = {\cal O}_{\g}
U_{\g,\g'}$ and $U_{\g,\g'}D_{\g'}\subset D_{\g}$ for all $g\ge g'$.}
in the appropriate sense. Hence, it follows from the general results
in \cite{5} that $[\E^3_i]_f(x)$, with domain ${\rm Cyl}^1$ (the space
of all $C^1$ cylindrical functions), is an essentially self-adjoint
operator on $\Ho$. For notational simplicity, we will denote its
self-adjoint extension also by $[\E^3_i]_f(x)$. (The context should
make it clear whether we are referring to the essentially self-adjoint
operator or its extension.)

Let us now turn to the integrand of the smeared area operator
(corresponding to (\ref{ra})). Denoting the determinant of the
intrinsic metric on $S$ by $g_S$, we have:
\ba\label{rg} 
[\hat{g}_S]_f (x)\,\cdot\, \Psi_\g &:=& [E^3_i]_f (x)
[E^{3i}]_f (x) \,\cdot\, \Psi_\g\nonumber\\ 
&=& -{\Pl^4\over 4}\big[ \sum_{I,J} \kappa(I,J) f_\e (x, v_{\alpha_I}) 
f_\e (x, v_{\alpha_J})\, X_I^i X_J^i\big] \,\cdot\, \Psi_\g\, ,
\ea
where the summation goes over all the oriented pairs $(I,J)$;
$v_{\a_I}$ and $v_{\a_J}$ are the vertices at which edges $e_I$ and
$e_J$ intersect $S$; $\kappa(I,J) = \kappa_I\kappa_J$ equals $0$ if
either of the two edges $e_I$ and $e_J$ fails to intersect $S$ or lies
entirely in $S$, $+1$ if they lie on the same side of $S$, and, $-1$
if they lie on the opposite sides. (For notational simplicity, from
now on we shall not keep track of the position of the internal indices
$i$; as noted in Sec. 2.3, they are contracted using the invariant
metric on the Lie algebra $su(2)$.) The next step is to consider
vertices $v_\a$ at which $\g$ intersects $S$ and simply rewrite the
above sum by re-grouping terms by vertices. The result simplifies if
we choose $\e$ sufficiently small so that, $f_\e(x, v_{\a_I}) f_\e(x,
v_{\a_J})$ is zero unless $v_{\a_I}= v_{\a_J}$.
We then have:
\be 
[\hat{g}_S]_f (x)\,\cdot\, \Psi_\g = - {\Pl^4\over 4}\big[\sum_{\a}\, 
 (f_\e(x, v_\a))^2\, \sum_{I_\a, J_\a} \kappa(I_\a, J_\a) 
X_{I_\a}^i X_{J_\a}^i\big]\,\cdot\, \Psi_\g\, ,
\ee
where the index $\a$ labels the vertices on $S$ and $I_\a$ and $J_\a$
label the edges at the vertex $\a$.

The next step is to take the square-root of this expression. The same
reasoning that established the self-adjointness of $[\E^3_i]_f(x)$ now
implies that $[\hat{g}_S]_f(x)$ is a non-negative self-adjoint
operator and hence has a well-defined square-root which is also a
positive definite self-adjoint operator. Since we have chosen $\e$
to be sufficiently small, for any given point $x$ in $S$, $f_\e
(x,v_\a)$ is non-zero for at most one vertex $v_\a$. We can therefore
take the sum over $\a$ outside the square-root. One then obtains
\be\label{rqae} 
([\hat{g}_S]_f)^{1\over 2} (x)\,\cdot\, \Psi_\g =  
{\Pl^2\over 2}\sum_{\a}\, f_\e(x, v_\a) \big[\sum_{I_\a, J_\a} 
\kappa(I_\a, J_\a) X_{I_\a}^i X_{J_\a}^i \big]^{1\over 2}\,\cdot\, 
\Psi_\g.
\ee
Note that the operator is neatly split; the $x$-dependence all resides
in $f_\e$ and the operator within the square-root is `internal' in
the sense that it acts only on copies of $SU(2)$.

Finally, we can remove the regulator, i.e., take the limit as
$\epsilon$ tends to zero. By integrating both sides against test
functions on $S$ and then taking the limit, we conclude that the
following equality holds in the distributional sense:
\be\label{qae} 
\widehat{\sqrt{{g}_S}} (x)\,\cdot\, \Psi_\g = {\Pl^2\over 2}\sum_{\a}\,
\delta^{(2)}(x, v_\a) \big[\sum_{I_\a, J_\a} \kappa(I_\a, J_\a)
X_{I_\a}^i X_{J_\a}^i \big]^{1\over 2}\cdot\, \Psi_\g.  
\ee
Hence, the regularized area operator is given by:
\be \label{AR} 
\hat{A}_S \,\cdot\, \Psi_\g = {\Pl^2\over 2} \sum_\a\,
\big[\sum_{I_\a, J_\a} \kappa(I_\a, J_\a) X_{I_\a}^i X_{J_\a}^i
\big]^{1\over 2}\,\cdot\, \Psi_\g .  
\ee
(Here, as before, $\a$ labels the vertices at which $\g$ intersects
$S$ and $I_\a$ labels the edges of $\g$ at the vertex $v_\a$.) With
${\rm Cyl}^2$ as its domain, $\hat{A}_S$ is essentially self-adjoint
on the Hilbert space $\Ho$.

The classical expression $A_S$ of (\ref{2.4}) is a rather complicated.
It is therefore somewhat surprising that the corresponding quantum
operators can be constructed rigorously and have quite manageable
expressions.  The essential reason is the underlying diffeomorphism
invariance which severely restricts the possible operators.  Given a
surface and a graph, the only diffeomorphism invariant entities are
the intersection vertices. Thus, a diffeomorphism covariant operator
can only involve structure at these vertices. In our case, it just
acts on the copies of $SU(2)$ associated with various edges at these
vertices.

We will close this discussion by simply writing the final expression of 
the volume operator:
\be
\hat{V}_R\cdot \Psi_\g := {\Pl^3\over 4\sqrt{3}}\, \sum_\a \,\, 
|\sum_{I_\a, J_\a, K_\a} i\, \e^{ijk}\, 
\e (I_\a, J_\a, K_\a)\, 
X^i_{I_\a}X^j_{J_\a} X^k_{K_\a}|^{1\over 2}\, 
\cdot \Psi_\g\, ,
\ee
where the first sum now is over vertices which lie in the region $R$
and $\epsilon(I_\a, J_\a, K_\a)$ is $0$ if the three edges are
linearly dependent at the vertex $v_\a$ and otherwise $\pm 1$
depending on the orientation they define. With $\Cyl^3$ as its domain,
$\hat{V}_R$ is essentially self-adjoint on $\Ho$.

To summarize, the diffeomorphism covariant functional calculus has
enabled us to regulate the area and volume operators. While in the
intermediate steps we have used additional structures --such as
charts-- the final results make no reference to these structures; the
final expressions of the quantum operators have the same covariance
properties as those of their classical counterparts.

\end{subsection}
\begin{subsection}{General properties of Geometric operators}
\label{s3.2}

We will now discuss the key properties of these geometric operators
and point out a few subtleties.  As in the previous subsection, for
definiteness, the detailed comments will refer to the area
operators. It should be clear from the discussion that analogous
remarks hold for the volume operators as well.
\medskip

1. {\em Discreteness of the spectrum:} By inspection, it follows that
the total area operator $\hat{A}_S$ leaves the sub-space of
$\Cyl^2_\g$ which is associated with any one graph $\g$ invariant and
is a self-adjoint operator on the sub-space $\Ho_\g$ of $\Ho$
corresponding to $\g$. Next, recall that $\Ho_\g = L^2(\A_\g,
d\mu^o)$, where $\A_\g$ is a compact manifold, isomorphic with
$(SU(2))^N$ where $N$ is the total number of edges in $\g$.  The
restriction of $\hat{A}_S$ to $\Ho_\g$ is given by certain commuting
elliptic differential operators on this compact manifold. Therefore,
all its eigenvalues are discrete. Now suppose that the complete
spectrum of $\hat{A}_S$ on $\Ho$ has a continuous part. Denote by
$P_c$ the associated projector. Then, given any $\Psi$ in $\Ho$,
$P_c\cdot\Psi$ is orthogonal to $\Ho_\g$ for any graph $\g$, and hence
to the space $\Cyl$ of cylindrical functions. Since $\Cyl^2$ is
dense in $\Ho$, $P_c\cdot\Psi$ must vanish for all $\Psi$ in
$\Ho$. Hence, the spectrum of $\hat{A}_S$ has no continuous part.

Note that this method is rather general: It can be used to show that
{\it any} self-adjoint operator on $\Ho$ which maps (the intersection
of its domain with) $\Ho_\g$ to $\Ho_\g$, and whose action on $\Ho_\g$
is given by elliptic differential operators, has a purely discrete
spectrum on $\Ho$. Geometrical operators, constructed purely from the
triad field tend to satisfy these properties.

In the case of area operators, one can do more: complete spectrum has
been calculated. The eigenvalues are given by \cite{al5}: 
\be\label{4.9} 
a_S = {\Pl^2\over 2}\, \sum_{\alpha}
\Big[2\jd_\alpha(\jd_\alpha+1) + 2\ju_\alpha(\ju_\alpha+1) -
\jdu_\alpha (\jdu_\alpha+1)\Big]^{1\over 2} 
\ee 
where $\a$ labels a finite set of points in $S$ and the non-negative
half-integers assigned to each $\a$ are subject to the inequality 
\be\label{4.8} 
\jd + \ju \ge \jdu \ge |\jd - \ju|\, .  
\ee 

There is, in particular, the smallest, non-zero eigenvalue, the
`elementary quantum of area': $a_S^o = (\sqrt{3}/4)\Pl^2$. Note,
however, that the level spacing between eigenvalues is {\it not}
regular. For large $a_S$, the difference between consecutive
eigenvalues in fact goes to zero as $1/\sqrt{a_S}$.  (For comparison
with other results \cite{9,10}, see \cite{al5}.)

2. {\em Area element:} Note that not only is the total area operator
well-defined, but in fact it arises from a local area element,
$\widehat{\sqrt{g_S}}$, which is an operator-valued distribution in the
usual sense. Thus, if we integrate it against test functions, the
operator is densely defined on $\Ho$ (with $C^2$ cylindrical functions
as domain) and the matrix elements
\be 
\langle {\Psi'}_{\g'},\, \widehat{\sqrt{g_S}}(x) \,\cdot\, 
\Psi_\g\rangle \nonumber
\ee
are 2-dimensional distributions on $S$. Furthermore, since we did not
have to renormalize the regularized operator (\ref{rqae}) before
removing the regulator, there are {\it no} free renormalization
constants involved. The local operator is completely unambiguous.

3. {\em $[\hat{g}_S]_f$ versus its square-root:} Although the
regulated operator $[\hat{g}_s]_f$ is well-defined, if we let
$\epsilon$ to go zero, the resulting operator is in fact divergent:
roughly, it would lead to the square of the 2-dimensional $\delta$
distribution. Thus, the determinant of the 2-metric is not a
well-defined in the quantum theory. As we saw, however, the
square-root of the determinant {\em is well} defined: We have to first
take the square-root of the {\em regulated} expression and {\it then}
remove the regulator. This, in effect, is the essence of the
regularization procedure.

To get around this divergence of $\hat{g}_S$, as is common in
Minkowskian field theories, we could have first rescaled
$[\hat{g}_S]_f]$ by an appropriate factor and then taken the
limit. Then result can be a well-defined operator, but it will depend
on the choice of the regulator, i.e., additional structure introduced
in the procedure. Indeed, if the resulting operator is to have the
same density character as its classical analog $g_S(x)$ --which is a
scalar density of weight two-- then the operator can not respect the
underlying diffeomorphism invariance.  For, there is no metric/chart
independent distribution on $S$ of density weight two. Hence, such a
`renormalized' operator is not useful to a fully non-perturbative
approach. For the square-root, on the other hand, we need a local
density of weight {\it one}. And, the 2-dimensional Dirac distribution
provides this; now is no apriori obstruction for a satisfactory
operator corresponding to the area element to exist. This is an
illustration of what appears to be typical in non-perturbative
approaches to quantum gravity: Either the limit of the operator exists
as the regulator is removed without the need of renormalization or it
inherits back-ground dependent renormalization fields (rather than
constants).

4. {\em Gauge invariance:} The classical area element $\sqrt{g_S}$ is
invariant under the internal rotations of triads $E^a_i$; its Poisson
bracket with the Gauss constraint functional vanishes.  This symmetry
is preserved in the quantum theory: the quantum operator
$\widehat{\sqrt{g_S}}$ commutes with the induced action of $\Gb$ on
the Hilbert space $\Ho$. Thus, $\widehat{\sqrt{g_S}}$ and the total
area operator $\hat{A}_S$ map the space of gauge invariant states to
itself; they project down to the Hilbert space $\tilde\Ho$ of
kinematic states. In the classical theory, the allowed values of the
area operators on the full phase space are the same as those on the
constraint surface. That is, the passage from all kinematical states
to the dynamically permissible ones does not give rise to restrictions
on the `classical spectrum' of these operators. The same is true in
the quantum theory. The spectrum of $\hat{A}_S$ on $\tilde\Ho$ is the
same as that on $\Ho$. (Only the degeneracies of eigenvectors
changes.)

\end{subsection}
\end{section}

\begin{section}{Discussion}
\label{s4}

In section 1, we began by formulating what we mean by quantization of
geometry: Are there geometrical observables which assume continuous
values on the classical phase space but whose quantum analogs have
discrete spectra? In order to explore these issues, we had to use a
fully non-perturbative framework which does not use a background
geometry.  In the last two sections, we answered the question in the
affirmative in the case of area and volume operators.  The
discreteness came about because, at the microscopic level, geometry
has a distributional character with 1-dimensional excitations. This is
the case even in semi-classical states which approximate classical
geometries macroscopically \cite{9,18}. We wish to emphasize that
these results have been obtained in the framework of a (non-traditional
but) rigorous quantum field theory. In particular, the issues of
regularization have been addressed squarely and the calculations are
free of hidden infinities.

We will conclude by examining the main results from various angles.

{\sl 1. Inputs}: The picture of quantum geometry that has emerged here
is strikingly different from the one in perturbative, Fock
quantization. Let us begin by recalling the essential ingredients that
led us to the new picture.

This task is made simpler by the fact that the new functional calculus
provides the degree of control necessary to distill the key
assumptions. There are only two essential inputs. The first assumption
is that the Wilson loop variables, $T_\a = {\rm Tr}\, {\cal P} \exp
\int_-\a A$, should serve as the configuration variables of the theory,
i.e., that the Hilbert space of (kinematic) quantum states should
carry a representation of the $C^\star$-algebra generated by the
Wilson loop functionals on the classical configuration space
$\ag$. The second assumption singles out the measure $\tilde\mu^o$.
In essence, if we assume that $\hat{E}^a_i$ be represented by $-i\hbar
{\delta/\delta A_a^i}$, the `reality conditions' lead us to the
measure $\tilde{\mu}^o$ \cite{7}.  Both these assumptions seem natural
from a mathematical physics perspective. However, a deeper
understanding of their {\it physical} meaning is still needed for a
better understanding of the overall situation.%
\footnote{In particular, in the standard spin-2 Fock representation, one
uses quite a different algebra of configuration variables and uses the
flat background metric to represent it. It then turns out that the
Wilson loops are {\it not} represented by well-defined operators; our
first assumption is violated. One can argue that in a fully
non-perturbative context, one can not mimic the Fock space
strategy. Further work is needed, however, to make this argument
water-tight.} %

{\sl 2. Kinematics versus Dynamics:} As was emphasized in the main
text, in the classical theory, geometrical observables are defined as
functionals on the {\it full} phase space; these are kinematical
quantities whose definitions are quite insensitive to the precise
nature of dynamics, presence of matter fields, etc.  Thus, in the
connection dynamics description, all one needs is the presence of a
canonically conjugate pair consisting of a connection and a (density
weighted) triad.  Therefore, one would expect the result on the area
operator presented here to be quite robust.  In particular, they
should continue to hold if we bring in matter fields or extend the
theory to supergravity.  

There is, however, a subtle caveat: In field theory, one can not
completely separate kinematics and dynamics.  For instance, in
Minkowskian field theories, the kinematic field algebra typically
admits an infinite number of {\it inequivalent} representations and a
given Hamiltonian may not be meaningful on a given representation.
Therefore, whether the kinematical results obtained in any one
representation actually hold in the physical theory depends on whether
that representation supports the Hamiltonian of the model. In the
present case, therefore, a key question is whether the quantum
constraints of the theory can be imposed meaningfully on $\tilde\Ho$.%
\footnote{Note that this issue arises in {\it any} representation once
a sufficient degree of precision is reached. In geometrodynamics, this
issue is not discussed simply because generally the discussion is
rather formal.} %
Results to date indicate (but do not yet conclusively prove) that this
is likely to be the case for general relativity.  The general
expectation is that this would be the case also for a class of
theories such as supergravity, which are `near' general relativity.
The results obtained here would continue to be applicable for this
class of theories.

{\sl 3. Dirac Observable:} Note that $\hat{A}_S$ has been defined for
{\it any} surface $S$.  Therefore, these operators will not commute
with constraints; they are not Dirac observables.  To obtain a Dirac
observable, one would have to specify $S$ {\it intrinsically}, using,
for example, matter fields.  In view of the Hamiltonian constraint,
the problem of providing an explicit specification is extremely
difficult.  However, this is true already in the classical theory.  In
spite of this, {\it in practice} we do manage to specify surfaces (or
3-dimensional regions) and furthermore compute their areas (volumes)
using the standard formula from Riemannian geometry which is quite
insensitive to the details of how the surface (region) was actually
defined.  Similarly, in the quantum theory, {\it if} we could specify
a surface $S$ (region $R$) intrinsically, we could compute the
spectrum of $\hat{A}_S$ and $\hat{V}_R$ using results obtained in this
paper.

{\sl 4. Manifold versus Geometry:} In this paper, we began with an
orientable, analytic, 3-manifold $\S$ and this structure survives in
the final description. As noted in footnote 1, we believe that the
assumption of analyticity can be weakened without changing the
qualitative results.  Nonetheless, a smoothness structure of the
underlying manifold will persist. What is quantized is `geometry' and
not smoothness. Now, in 2+1 dimensions, using the loop representation
one can recast the final description in a purely combinatorial fashion
(at least in the so-called `time-like sector' of the theory). In this
description, at a fundamental level, one can avoid all references to
the underlying manifold and work with certain abstract groups which,
later on, turn out to be the homotopy groups of the
`reconstructed/derived' 2-manifold (see, e.g., section 3 in
\cite{2+1}). One might imagine that if and when our understanding of
knot theory becomes sufficiently mature, one would also be able to get
rid of the underlying manifold in the 3+1 theory and introduce it
later as a secondary/derived concept. At present, however, we are
quite far from achieving this.

In the context of geometry, however, a detailed combinatorial picture
{\it is} emerging. Geometrical quantities are being computed by
counting; integrals for areas and volumes are being reduced to genuine
sums. (However, the sums are {\it not} the `obvious' ones, often used
in approaches that {\it begin} by postulating underlying discrete
structures. In the computation of area, for example, one does not just
count the number of intersections; there are precise and rather
intricate algebraic factors that depend on the representations of
$SU(2)$ associated with the edges at each intersection.) It is
striking to note that, in the same address \cite{Rie} in which Riemann
first raised the possibility that geometry of space may be a physical
entity, he also introduced ideas on discrete geometry. The current
program comes surprisingly close to providing us with a concrete
realization of these ideas.
\medskip

To summarize, it {\it is} possible to do physics in absence of a
background space-time geometry. It does require the use of new
mathematical methods, such as a diffeomorphism covariant functional
calculus. However, one can obtain concrete, physically motivated
results which are quite surprising from the viewpoint of Minkowskian
field theories.

\end{section}

\bigskip
{\centerline {\bf Acknowledgments}} 

We would also like to thank John Baez, Bernd Bruegman, Don Marolf,
Jose Mourao, Thomas Thiemann, Lee Smolin, John Stachel and especially
Carlo Rovelli for discussions.  This work was supported in part by the
NSF Grants PHY93-96246 and PHY95-14240, the KBN grant 2-P302 11207 and
by the Eberly fund of the Pennsylvania State University. JL thanks the
members of the Max Planck Institute for their hospitality.


\begin{thebibliography}{99}
\bibitem{cr} C. Rovelli, How much of quantum field theory is compatible 
with what we know about spacetime?
\end{thebibliography}
\end{document}